\documentclass[a4paper]{report}
\usepackage[utf8]{inputenc}
\usepackage[T1]{fontenc}
\usepackage{RJournal}
\usepackage{amsmath,amssymb,array}
\usepackage{booktabs}

\usepackage{float}
\floatstyle{ruled}
\newfloat{Algorithm}{htbp}{alg}

\DeclareMathOperator{\sech}{sech}

\begin{document}

\sectionhead{Contributed research article}
\volume{XX}
\volnumber{YY}
\year{20ZZ}
\month{AAAA}

\begin{article}

\title{gk: An R Package for the $g$-and-$k$ and generalised $g$-and-$h$ Distributions}
\author{by Dennis Prangle}

\maketitle

\abstract{ 
The $g$-and-$k$ and (generalised) $g$-and-$h$ distributions are flexible univariate distributions which can model highly skewed or heavy tailed data through only four parameters:
location and scale, and two shape parameters influencing the skewness and kurtosis.
These distributions have the unusual property that they are defined through their quantile function (inverse cumulative distribution function) and their density is unavailable in closed form, which makes parameter inference complicated.
This paper presents the \pkg{gk} R package to work with these distributions.
It provides the usual distribution functions
and several algorithms for inference of independent identically distributed data,
including the finite difference stochastic approximation method, which has not been used before for this problem.
}

\section{Introduction}

Statisticians have long sought for a simple extension to the normal distribution which can model data subject to skew, heavy tails or both.
One approach is to transform a standard normal random variable $Z \sim N(0,1)$ to
\begin{equation} \label{eq:Qgen}
X = A + B G(Z) H(Z),
\end{equation}
where $A$ and $B$ are location and scale parameters,
$G(\cdot)$ introduces asymmetry,
and $H(\cdot)$ elongates the tails of the distribution while having little effect near the mode.
This paper considers two such distributions, the $g$-and-$k$ and generalised $g$-and-$h$ distributions.
These distributions can model many types of behaviour through just a small number of parameters.

Defining random variables as transformations of $Z$ is equivalent to specifying the distribution's quantile function (defined in the next section),
and distributions of this type are known as \dfn{quantile distributions}.
Work on quantile distributions goes back at least to \citet{Hastings:1947}.
See \citet{Gilchrist:2000} for a book length treatment of their history and use in statistics.
\citet{Tukey:1977} proposed the form \eqref{eq:Qgen} and a distribution using it: the original $g$-and-$h$ distribution.
\citet{Haynes:1997} were the first to use the two distributions considered in this paper: the $g$-and-$k$ distribution and a generalised form of the $g$-and-$h$ distribution.
For brevity henceforth ``$g$-and-$h$ distribution'' will refer to their generalised form.
See \cite{Peters:2016} for a thorough review of these and other distributions based on \eqref{eq:Qgen}.

Applications of the $g$-and-$k$ and $g$-and-$h$ distributions have included environmental data \citep{Rayner:2002}, financial returns \citep{Drovandi:2011} and insurance risk \citep{Peters:2016}.
There has also been considerable methodological work on inference for these distributions \citep[e.g.][]{Rayner:2002, Haynes:2005, Allingham:2009, Drovandi:2011, Fearnhead:2012}.
This is because it is not possible to express the densities of quantile distributions in closed form beyond some special cases, which makes it difficult to apply standard likelihood-based inference methods.

This paper presents the \pkg{gk} R package to work with the $g$-and-$k$ and $g$-and-$h$ distributions.
The remaining sections covering the following:
\begin{itemize}
\item A mathematical definition of the distributions.
\item A description of the package's functions to perform standard distributional tasks and how they are implemented.
\item An exploration of the range of valid parameters for these distributions, as this has a complicated form. We propose a novel rule giving ``safe'' parameter values for the $g$-and-$k$ distribution.
\item A desctiption of several methods for parameter inference and corresponding functions supplied by the package.
\item An illustrative analysis of a real dataset.
\item A summary.
\end{itemize}

\section{Definitions}

The cumulative distribution function (cdf) of a univariate random variable $X$, $F_X:\mathbb{R} \to [0,1]$, is defined as $\Pr(X \leq x)$.
(Later we will often drop subscripts where they are clear from the context.)
The cdf suffices to completely specify the probability distribution of $X$.
It is often the case that the cdf is not available in closed form but is implicitly defined through its derivative, the probability density function (pdf).
An example of this is the normal distribution.

For quantile distributions, the cdf is implicity defined through its inverse, the \dfn{quantile function} $F_X^{-1}(u)$ where $F_X^{-1}:[0,1] \to \mathbb{R}$.
The $g$-and-$k$ and $g$-and-$h$ distributions use a quantile function of the form
$F^{-1}(u;\theta) = Q(z(u);\theta)$ where $z(\cdot)$ is the $N(0,1)$ quantile function and $\theta$ is a vector of parameters.
The $Q$ functions are:
\begin{align}
Q_{gk}(z;A,B,g,k,c) &= A + B(1+c \tanh[gz/2])z (1+z^2)^k \label{eq:Qgk} \\
Q_{gh}(z;A,B,g,h,c) &= A + B(1+c \tanh[gz/2])z \exp(hz^2/2). \label{eq:Qgh}
\end{align}
It is possible to sample from the distributions using the \dfn{inversion method},
that is, by simulating $U \sim U(0,1)$ and substituting it into the quantile function.
Equivalently one can sample $Z \sim N(0,1)$ and substitute it into $Q_{gk}$ or $Q_{gh}$
i.e.~the process described in the introduction based on Equation \eqref{eq:Qgen}.
In terms of \eqref{eq:Qgen},
$G(z) = 1+c \tanh(gz/2)$ produces asymmetry and
$H(z) = z(1+z^2)^k$ or $z \exp(hz^2/2)$ elongates tails.

\begin{figure}[htp]
\includegraphics[width=\textwidth]{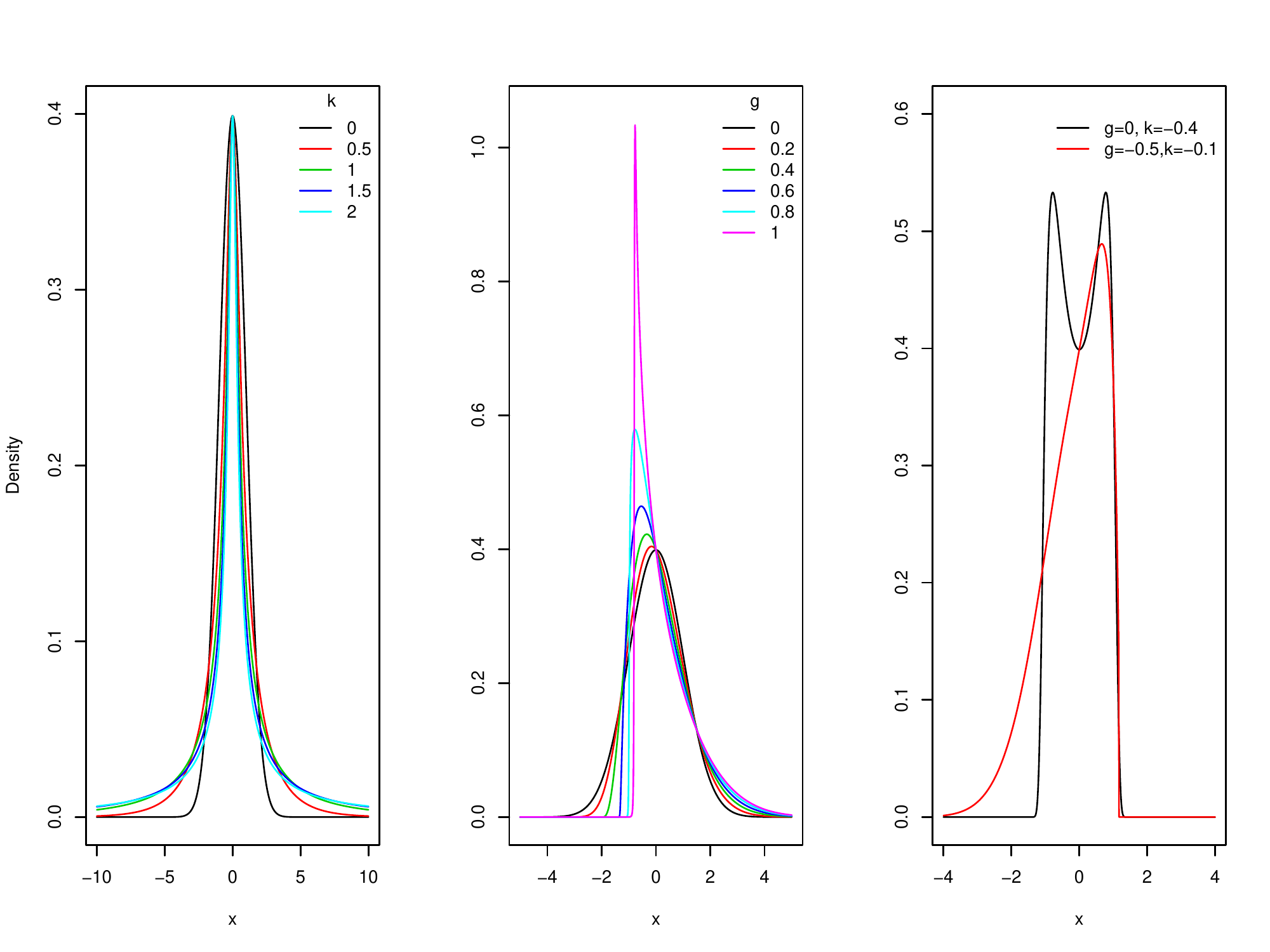}
\caption{Example $g$-and-$k$ densities. The first panel fixes $g=0$ and varies $k$, mainly altering kurtosis. The second fixes $k=0$ and varies $g$, mainly altering skewness. The third shows two examples with $k<0$.}
\label{fig:gk_densities}
\end{figure}

Each distribution has four main parameters:
$A$ (location),
$B$ (scale),
$g$ (shape parameter mainly affecting skewness),
and $k$ or $h$ (shape parameter mainly affecting kurtosis).
The remaining parameter $c$ is discussed below.
When both shape parameters are zero the distribution is simply $N(0,1)$.
An illustration of the flexible shapes that the $g$-and-$k$ density can take is given in Figure \ref{fig:gk_densities}.
The $g$-and-$h$ can produce similar shapes, with the following exception.
The $g$-and-$k$ distribution allows negative values of $k$ which can produce lighter tails than a normal distribution, but also bimodal distributions of potentially limited usefulness.

Well-defined continuous distributions result from parameter values producing strictly increasing quantile functions.
Determining when this is true is complicated so discussion is postponed to a later section.
For now note that it is standard to take $B>0$ and fix $c=0.8$ (which will be assumed throughout unless mentioned otherwise), and in this case $k \geq 0$ or $h \geq 0$ guarantees a valid distribution.

\section{Distribution functions}

The \pkg{gk} package provides the standard suite of R functions for the $g$-and-$k$ and $g$-and-$h$ distributions
i.e.~random sampling and calculation of the pdf, cdf and quantile functions.
This section describes how these functions are implemented.
It is assumed that parameters have been chosen such that the quantile function is strictly increasing.
No warning is given when this is not the case as checking validity is time consuming (see next section).

\paragraph{Quantile function}

The \code{qgk} and \code{qgh} functions calculate the quantile function $F^{-1}(u)$.
Their implementation is straightforward.
First $z(u)$ is calculated using \code{qnorm},
then this is passed to an internal function, \code{z2gk} or \code{z2gh}, which computes $Q_{gk}$ or $Q_{gh}$.

\paragraph{Random sampling}

The \code{rgk} and \code{rgh} functions perform random sampling.
This is done by the method described earlier of sampling $N(0,1)$ draws and substituting them into $Q_{gk}$ or $Q_{gh}$, via the function \code{z2gk} or \code{z2gh}.

\paragraph{Cumulative distribution function}

The \code{pgk} and \code{pgh} functions calculate the cdf $F(x)$ given input $x$.
They numerically solve $Q(z) - x = 0$, which is guaranteed to have a unique root for $z$.
The required final output is then $u = \Phi^{-1}(z)$ where $\Phi$ is the $N(0,1)$ cdf.
An alternative approach would be to directly solve $Q(z(u)) - x = 0$ for $u$.
However we found this was less numerically stable for $u$ close to 0 or 1.

Our code finds the root for $z$ using R's \code{uniroot} command and \code{z2gk} or \code{z2gh} for $Q(z)$ evaluations.
The need to run a root finding algorithm means this function is slow relative to cdf calculations of standard distributions
- see Table \ref{tab:cost}.

The functions include an argument \code{zscale}.
Setting this to \code{TRUE} outputs the $z$ value which is found rather than $u$.
This is used in the density functions below,
and more generally is also useful to retain numerical precision when $z$ has large magnitude.

\paragraph{Probability density function}

The \code{dgk} and \code{dgh} functions calculate the pdf $f(x)$,
or the log pdf if the argument \code{log=TRUE} is supplied.
The method is based on the standard probability result that if $A$ has density $f_A(a)$ and $t(a)$ is a differentiable 1-1 transformation then the density of $B=t(A)$ is
\[
f_B(b) = f_A(a) / t'(a) \qquad \text{where } a = t^{-1}(b),
\]
and $t'$ denotes the first derivative of $t$.

For quantile distributions we have $Z \sim N(0,1)$ and $X = Q(Z)$ for some $Q$ function.
So the pdf of $X$ is
\[
f(x) = \phi(z)/Q'(z) \qquad \text{where } z = Q^{-1}(x),
\]
where $\phi(z)$ is the $N(0,1)$ pdf.

Our code to calculate the pdf first finds $z = Q^{-1}(x)$ using \code{pgk} or \code{pgh} with \code{zscale=TRUE}.
Then the pdf or its log is calculated using
formulae \eqref{eq:Qgk'} and \eqref{eq:Qgh'} (see Appendix A) for $Q'(u)$.
The reliance on performing root finding within \code{pgk} and \code{pgh} means that \code{dgk} and \code{dgh} are slow relative to pdf calculations for standard distributions
- see Table \ref{tab:cost}.

Note that an alternative representation of $f(x)$ is $1/q'(u)$ where $q(u)$ represents $F^{-1}(u)$ and $u = F(x)$.
Density calculations based on this approach are described in \citet{Rayner:2002}.
However we found that calculating the $u$ values required for this approach was occasionally numerically unstable, as mentioned above.

\paragraph{Cost}

Table \ref{tab:cost} compares the time to execute \pkg{gk}'s distributional functions to those for the normal distribution.
It illustrates that random sampling and quantile function calculation are reasonably efficient, but calculating the cdf and pdf are expensive.

\begin{table}[htp]
\centering
\begin{tabular}{rrrrrr}
  \hline
                  & \multicolumn{3}{c}{Time (microseconds)} & \multicolumn{2}{c}{Ratio vs normal} \\
                  & Normal & $g$-and-$k$ & $g$-and-$h$ & $g$-and-$k$ & $g$-and-$h$ \\ 
  \hline
Quantile function & 175    & 972         & 445         & 5.56        & 2.55        \\ 
Random sampling   & 150    & 921         & 436         & 6.15        & 2.91        \\ 
cdf               & 313    & 143151      & 116928      & 457         & 374         \\ 
pdf               & 369    & 138381      & 111279      & 375         & 302         \\ 
   \hline
\end{tabular}
\caption{Mean times to perform various distributional operations, evaluated by the \pkg{microbenchmark} pacakge.
For example the random sampling row compares \code{rnorm(N)}, \code{rgk(N,1,2,3,4)} and \code{rgh(N,1,2,3,4)} for $N=100$.
We also tried $N=1$, which gave qualitatively similar results but slightly better relative efficiency of the \pkg{gk} functions.}
\label{tab:cost}
\end{table}

\section{Range of valid parameters}

Recall that a valid continuous distribution requires the quantile function to be strictly increasing.
Clearly this property is unaffected by the choice of $A$ and $B>0$.
This section discusses the effects of $g,h,k$ and $c$.

Several theoretical results on valid parameters can be derived.
It's convenient to concentrate on $c \geq 0$.
In this case $h < 0$ or $k < -1/2$ is invalid.
Taking $k \geq 0$ or $h \geq 0$ produces valid distributions when $0 \leq c < c^* \approx 0.83$.
This is the reason for taking $c=0.8$ as standard: it maintains this property while allowing the skewness factor in \eqref{eq:Qgk} and \eqref{eq:Qgh} to have a large effect.
For justification of all these results, see Appendix B.

When $c=0.8$, the above results completely characterise the range of valid parameters for the $g$-and-$h$ distribution.
For the $g$-and-$k$ distribution, there is still some uncertainty for $-0.5 \leq k<0$,
which, as mentioned earlier, corresponds to light tails.
For both distributions, the case where $c>c^*$ is less clear: even positive values of $k$ or $h$ do not guarantee validity.
Therefore we provide the function \code{isValid} to test parameter validity numerically.

Validity can be checked by testing whether the minimum derivative of \eqref{eq:Qgk} or \eqref{eq:Qgh} is positive.
Appendix A shows that it is equivalent to test whether the functions \eqref{eq:Rgk} or \eqref{eq:Rgh} are positive.
\code{isValid} uses numerical optimisation to minimise these and returns whether the minimum value is positive.
To reduce the possibility of finding local minima, multiple optimisation starting points can supplied as a vector to the argument \code{initial\_z}.
However it is still not guaranteed that the global minimum is found, so there remains a possibility that the function may produce false positives.

\begin{figure}[htp]
\includegraphics[width=\textwidth]{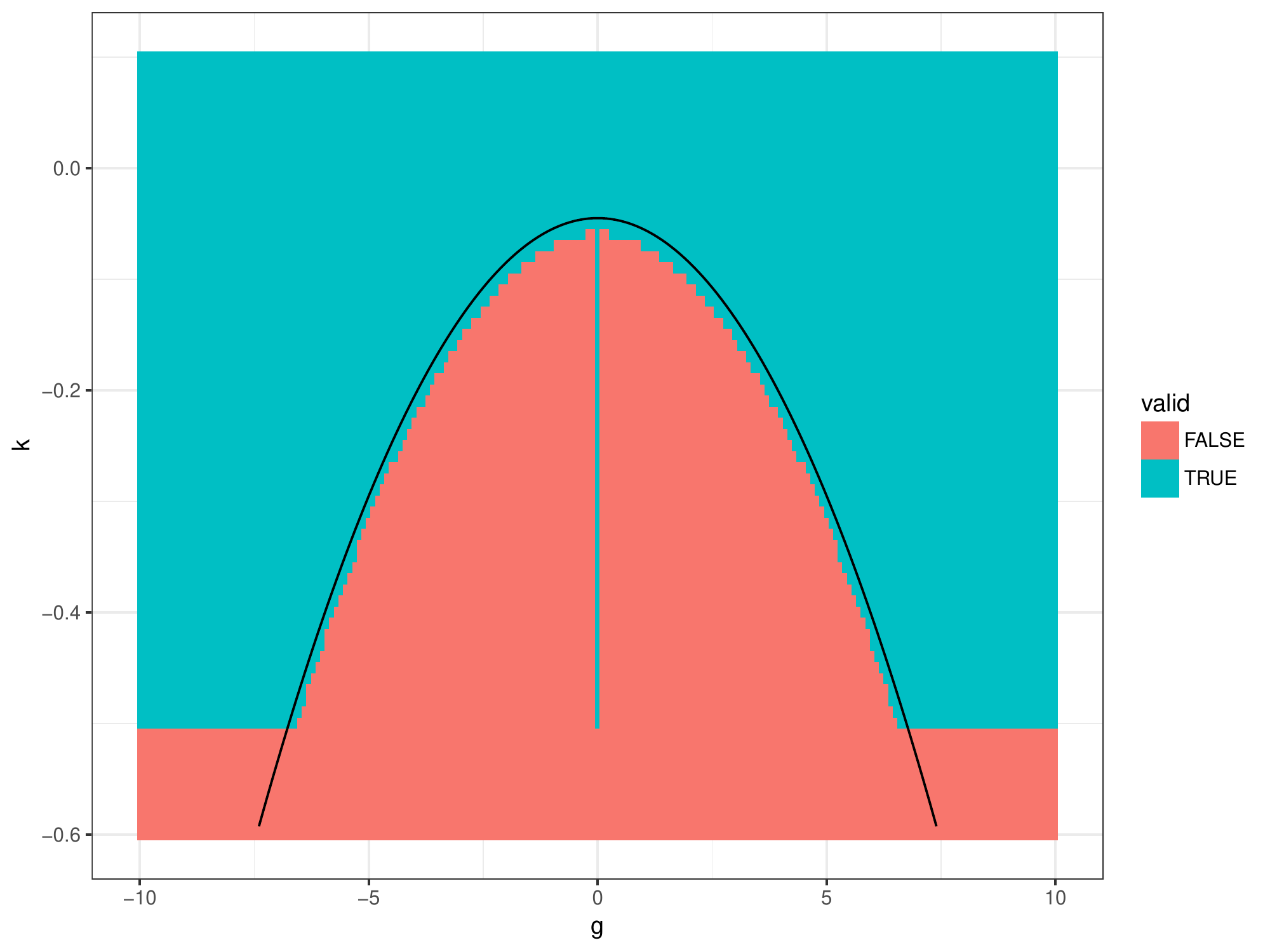}
\caption{Validity of parameter values for the $g$-and-$k$ distribution when $c=0.8$, calculated using \code{isValid}.
Also shown is a quadratic function $\tilde{k}(g)$ near the curved part of the boundary between the regions.}
\label{fig:gk_valid}
\end{figure}

The function can be used as follows to illustrate the region of valid $g$-and-$k$ parameter values for $c=0.8$.
The results are plotted as Figure \ref{fig:gk_valid}.

\begin{example}
gk_grid = expand.grid(g = seq(-10,10,0.1), k = seq(-0.6,0.1,0.01)) 
v = isValid(gk_grid$g, gk_grid$k)
\end{example}

We do not test validity automatically within the package's other functions.
This is because \code{isValid} is relatively computationally expensive and not guaranteed to be correct.
Therefore particular care should be taken for $k<0$ or $c > c^*$, as the distribution functions will not provide warnings when invalid parameters are used.
A reasonable region of $g$ and $k$ values to use in practice with $c=0.8$ can be derived from Figure \ref{fig:gk_valid}.
It shows that for $|g|<7$ some $-0.5 \leq k < 0$ values are invalid.
Apart from a narrow strip near $g=0$, this invalid region's boundary is roughly quadratic, as illustrated by the curve $\tilde{k}(g) = -0.045-0.01g^2$ on the figure.
Based on this analysis, $k \geq \max(-0.5, \tilde{k}(g))$ seems a reasonable sufficient condition for parameter validity to use in practice.

\section{Inference functions}

The package provides three inference methods for data $x_1,x_2,\ldots,x_n$
which are assumed to be independent and identically distributed (IID)
draws from a $g$-and-$k$ or $g$-and-$h$ distribution with unknown parameters.
This section describes these methods.
An illustration of their use is provided in the next section.
See the \pkg{gk} help files for a full description of all the arguments available.

\paragraph{MCMC inference}

The \code{mcmc} function implements inference using \dfn{Markov chain Monte Carlo} (MCMC).
This samples from a Markov chain whose stationary distribution is the Bayesian posterior of interest for the parameters $\theta$.
We use a Metropolis-Hastings algorithm, in which a proposed new state of the chain $\theta'$ is sampled by adding a $N(0,\Sigma)$ increment to the current state $\theta_t$.
A decision to accept or reject $\theta'$ is made based on the prior and likelihood values at $\theta_t$ and $\theta'$ and a random variable (see steps 3-4 of Algorithm \ref{alg:mcmc}.)

Tuning $\Sigma$ can be difficult.
\cite{Haynes:2005}, who first used MCMC for the $g$-and-$k$ distribution, did this manually.
Instead we use the adaptive Metropolis (AM) algorithm of \citet{Haario:2001} which tunes $\Sigma$ automatically during its operation.
The resulting $\theta_t$s no longer form a Markov chain, but it has been proved \citep{Saksman:2010} that, under suitable conditions, calculations using them still converge to posterior quantities as the length of the chain increases.
The AM algorithm is presented as Algorithm \ref{alg:mcmc}.
Step 1 states the proposal matrix used in terms of the empirical variance of the past MCMC states.
To calculate this empirical variance efficiently, the code updates it each time a new state is observed.
As a default we specify tuning choices $\epsilon = 10^{-6}$ and $t_0=100$.

Like other Bayesian methods, MCMC requires a prior density for the parameters, $\pi(\theta)$, to be specified.
This must be supplied by the user.
For computational convenience this should be supplied in the form of a function \code{get\_log\_prior} which takes a vector of parameters as input and returns the log prior density.
We allow the user to reparameterise $\theta$, using $\log B$ rather than $B$, via the \code{logB} argument.
This can improve MCMC efficiency when the posterior for $B$ is concentrated on values close to zero.

For IID data the likelihood is $L(\theta) = \prod_{i=1}^n f(x_i;\theta)$, the product each observation's pdf.
Evaluating this for the $g$-and-$k$ or $g$-and-$h$ distributions using the \code{pgk} or \code{pgh} command requires $n$ calls to numerical optimisation.
Therefore MCMC becomes computationally expensive for even moderately large datasets.

\begin{Algorithm} \caption{The Adaptive Metropolis MCMC algorithm} \label{alg:mcmc}
\begin{itemize}
\item[] \strong{Input}: observations $x$, prior density $\pi(\theta)$, number of iterations to perform $N$, initial state $\theta_0$, initial variance matrix $\Sigma_0$, pre-tuning period $t_0$, tuning parameter $\epsilon>0$.
\end{itemize}
\begin{enumerate}
\item[] \strong{Loop} over $1 \leq t \leq N$:
\item If $t \leq t_0$ let $\Sigma_t = \Sigma_0$. Otherwise let $\Sigma_t = \tfrac{1}{4}(2.4)^2 (\hat{\Sigma}_{t-1}+\epsilon I)$, where $\hat{\Sigma}_{t-1}$ is the variance of $\theta_1,\theta_2,\ldots,\theta_{t-1}$.
\item Sample $\theta' \sim N(\theta_{t-1}, \Sigma_{t-1})$
\item Sample $u \sim U(0,1)$ and let $r = \frac{\pi(\theta') L(\theta')}{\pi(\theta_{t-1}) L(\theta_{t-1})}$.
\item If $u<r$ let $\theta_t = \theta'$. Otherwise let $\theta_t = \theta_{t-1}$.
\end{enumerate}
\begin{itemize}
\item[] \strong{Output}: sample $\theta_0,\theta_1,\ldots,\theta_N$.
\end{itemize}
\end{Algorithm}

\paragraph{ABC inference}

The \code{abc} function implements inference by \dfn{approximate Bayesian computation} (ABC).
This is a method for approximate Bayesian inference which avoids evaluating the likelihood function.
It is especially useful when the likelihood function is unavailable or, as for quantile distributions, is expensive to compute.
ABC is based instead on finding parameter values which produce simulated data similar to the observations.
The \code{abc} function implements a simple version of ABC, Algorithm \ref{alg:ABC}.
Here a simulation is accepted if it has one of the $M$ smallest distances to the observations.
Distance refers to a weighted version of Euclidean distance between vectors of simulated and observed \dfn{summary statistics}.
Details of the weighting are given in the algorithm's description.
(For $n$ large, \code{abc} avoids high memory requirements by running several batches of Algorithm \ref{alg:ABC}.
Each batch uses $N=10^4$ and returns the $M$ best simulations.
The overall best $M$ best simulations are then found and returned.
The $v_j$ weights calculated in the first batch are reused in the others.)

Like \code{mcmc}, \code{abc} is a Bayesian method and requires a prior distribution for $\theta$ to be provided.
It is convenient for this to be provided in a different form to the \code{mcmc} case.
A function \code{rprior} should be supplied which a single numeric input and returns that many samples from the prior distribution as rows of a matrix.

\begin{Algorithm} \caption{Approximate Bayesian computation (ABC)} \label{alg:ABC}
\begin{itemize}
\item[] \strong{Input}: observations $x$, prior distribution $\pi(\theta)$, summary statistic function $s(x)$, number of simulations to perform $N$, number of samples to output $M$.
\end{itemize}
\begin{enumerate}
\item Calculate observed summaries $s_0=s(x)$.
\item For $1 \leq i \leq N$ sample parameters $\theta_i$ from the prior.
\item For $1 \leq i \leq N$ simulate summary statistics $s(x_i)$ given parameters $\theta_i$. Let $s_{ij}$ denote the $j$th component of $s(x_i)$.
\item For $1 \leq j \leq q$ (where $q=\dim(s_0)$) calculate the empirical variance $v_j$ of the $(s_{ij})_{1 \leq i \leq n}$ values.
\item For $1 \leq i \leq N$ let $d_i = \sum_{j=1}^q (s_{ij}-s_{0j})^2/v_j$.
\item Find the $M$ smallest $d_i$ values and return the corresponding $\theta_i$s.
\end{enumerate}
\end{Algorithm}

ABC produces samples from an approximation to the Bayesian posterior distribution.
The quality of the approximation depends in a complex way on the choice of summary statistics and the tuning parameters $N$ and $M$.
For more background on ABC see the review paper by \cite{Marin:2012} and the handbook of \cite{Sisson:2017}.
Two general R packages for ABC which implement more advanced methods are \pkg{abc} \citep{Csillery:2012} and \pkg{easyABC} \citep{Jabot:2013}.

Using ABC for the $g$-and-$k$ and $g$-and-$h$ distributions was proposed by \cite{Allingham:2009} and has been investigated in many subsequent papers.
Following \cite{Drovandi:2011} we offer three choices of summary statistics which can be selected through the \code{sumstats} argument:
(1) the full order statistics;
(2) octiles of the observations, $E_1,E_2,\ldots,E_7$;
(3) robust estimates of the moments based on the octiles:
\[
S_A=E_4,\quad S_B=E_6-E_2,\quad S_g=(E_6+E_2-2E_4)/S_b,\quad S_k=(E_7-E_5+E_3-E_1)/S_b.
\]
Many more sophisticated approaches to choosing ABC summary statistics have been proposed \citep{Blum:2013}, but these are a simple starting point.

For summaries (2) or (3) we follow \cite{Fearnhead:2012} and speed up step 3 of Algorithm \ref{alg:ABC} by using the fact that the octiles (or close approximations) can be simulated quickly without the need to simulate a full dataset.
Suppose $X_1,X_2,\ldots,X_N$ are $g$-and-$k$ or $g$-and-$h$ variables,
and let $X_{(1)} < X_{(2)} \ldots < X_{(N)}$ denote the order statistics.
We replace $E_i$ with $E'_i = X_{(r(iN/8))}$ where $r(\cdot)$ rounds to the nearest integer.
Now we need to simulate 7 order statistics from the $g$-and-$k$ or $g$-and-$h$ distribution.
To do so we simulate corresponding order statistics of the $\text{Uniform}(0,1)$ distribution using the exponential spacings method \citep{Ripley:1987}.
This is implemented by the \code{orderstats} function.
The uniform order statistics are then substituted into $F^{-1}(u)$.

\paragraph{FDSA inference}

The \code{fdsa} function performs inference using \dfn{finite difference stochastic approximation} (FDSA).
FDSA, originally due to \citet{Kiefer:1952}, attempts to find $\theta^*$ minimising a loss function $\mathcal{L}(\theta)$ by iteratively calculating estimates $\theta_1, \theta_2, \ldots$.
Each iteration moves the estimate in the opposite direction to an estimate of the loss gradient, based on finite difference calculations.

We use FDSA for maximum likelihood estimation of IID observations.
In this setting $\mathcal{L}(\theta)$ can be taken to be the negative log likelihood,
\[
\mathcal{L}(\theta) = -\log L(\theta) = -\sum_{i=1}^n \log f(x_i;\theta).
\]
The gradient of $\mathcal{L}(\theta)$ can be estimated using only a small subset of the data, so FDSA has the potential to scale up to large datasets better than MCMC, while avoiding the approximation error of ABC.
Unlike ABC and MCMC, we are not aware of FDSA having previously been used for the $g$-and-$k$ and $g$-and-$h$ distributions.

The $g$-and-$k$ and $g$-and-$h$ distributions have some parameter constaints (e.g.~$B>0$, $h \geq 0$).
Also we found setting further constaints from preliminary analyses sometimes helps FDSA behave well.
Therefore we use a version of FDSA for bounded minimisation from \citet{L'Ecuyer:1994},
presented as Algorithm \ref{alg:FDSA}.

\begin{Algorithm} \caption{Finite difference stochastic approximation (FDSA)} \label{alg:FDSA}
\begin{itemize}
\item[] \strong{Input}: initial state $\theta_0$, choice of $a_t$ and $c_t$ sequences, function $\hat{\mathcal{L}}(\cdot)$ which calculates an unbiased estimate of $\mathcal{L}(\cdot)$, number of iterations to perform $N$, vectors of (possibly infinite) upper and lower parameter bounds $\theta^+, \theta^-$.
\end{itemize}
\begin{enumerate}
\item[] \strong{Loop} over $0 \leq t \leq N-1$.
\item Calculate $\hat{g}_t$ by performing the following steps for $i \leq 1 \leq 4$.
\begin{enumerate}
\item Let $\Delta_i$ be a $4$-dimensional vector whose $i$th component is 1 and others are zero.
\item Let $\phi^+ = P(\theta_t + c_t \Delta_i)$ and $\phi^- = P(\theta_t - c_t \Delta_i)$. \\
(Here $P(\phi)$ is a projection operator.
Its output is $\phi'$ such that $\phi'_i$ is the closest value to $\phi_i$ in $[\theta^-_i, \theta^+_i]$.
The $i$ subscripts represent $i$th components.)
\item Let $\hat{g}_{it} = \frac{1}{|\phi^+_i-\phi^-_i|}[\hat{\mathcal{L}}(\phi^+) - \hat{\mathcal{L}}(\phi^-)]$.
\end{enumerate}
\item Let $\theta_{t+1} = P(\theta_t - a_t \hat{g}_t)$.
\item[] \strong{Output:} Final estimate $\theta_N$.
\end{enumerate}
\end{Algorithm}

The unbiased estimate of $\mathcal{L}(\theta)$ required by Algorithm \ref{alg:FDSA}, $\hat{\mathcal{L}}(\theta)$, can be taken to be the sum of a random sample of $m$ negative log likelihood terms multiplied by $n/m$.
Hence for a vector \code{y} containing a random subsample of $m$ observations
(sometimes referred to as a \emph{batch}),
$\hat{\mathcal{L}}(\theta)$ can be calculated using \code{-sum(dgk(y,A,B,g,k,log=TRUE))*n/m} (or similar for the $g$-and-$h$ distribution).
Variance reduction in step 1c of Algorithm \ref{alg:FDSA} is possible by coupling the two estimates \citep{Kushner:2003}.
Hence we use the same random subsample of data for all $\hat{\mathcal{L}}$ calculations in an iteration of step 1.

FDSA convergence requires that the \dfn{gain sequences} $a_t$ and $c_t$ must satisfy certain conditions.
Following \cite{Spall:1998} we take $a_t = a_0 (A+t+1)^{-\alpha}$ and $c_t = c_0 (t+1)^{-\gamma}$.
This leaves several tuning choices, which can be selected by the user, or left at default values which we provide.
Following \cite{Kleinman:1999} we use default values $\alpha=1$ and $\gamma=0.49$.
Following \cite{Spall:1998} our default for $c_0$ is an estimate of the standard deviation of $\hat{\mathcal{L}}(\theta_0)$ using some preliminary simulations.
We provide defaults $a_0=1$ and $A=100$ but it is recommended to manually tune these to produce rapid convergence.
This may require several short pilot runs of the algorithm.
The \code{fdsa} function allows $a_0$ and $c_0$ to be vectors, in which case operations in Algorithm \ref{alg:FDSA} are interpreted as elementwise where necessary.
This allows the user to tune gain sequences differently for each parameter.
As for \code{mcmc}, we allow the user to reparameterise $\theta$, using $\log B$ rather than $B$, via the \code{logB} argument, which can improve FDSA efficiency when the MLE value of $B$ is close to zero.

Under weak assumptions, FDSA converges to a local minimum of $\mathcal{L}(\theta)$ \citep{Kushner:2003}.
In our experience the likelihood for the $g$-and-$k$ and $g$-and-$h$ distributions is usually unimodal, so there is little danger of converging to an incorrect mode.
Nonetheless it may be a useful check on the results to rerun the algorithm from various starting points or compare with the output of another algorithm.

An alternative to FDSA is \dfn{simultaneous perturbation stochastic approximation} (SPSA) \citep{Spall:1998}.
Here each iteration makes a finite difference estimate of the derivative of the loss function when moving in a random direction from $\theta_t$.
An update moves $\theta_t$ a distance (negatively) proportional to this estimated derivative in the selected direction.
Each SPSA iteration requires fewer likelihood estimates than FDSA, and it is asymptotically more efficient \citep{Kushner:2003}.
However we found in exploratory work that for our application the SPSA updates were dominated by improving $A$ and $B$ estimates, and the remaining parameters were learned very slowly.

\section{Illustration}

\begin{figure}[htp]
\includegraphics[width=\textwidth]{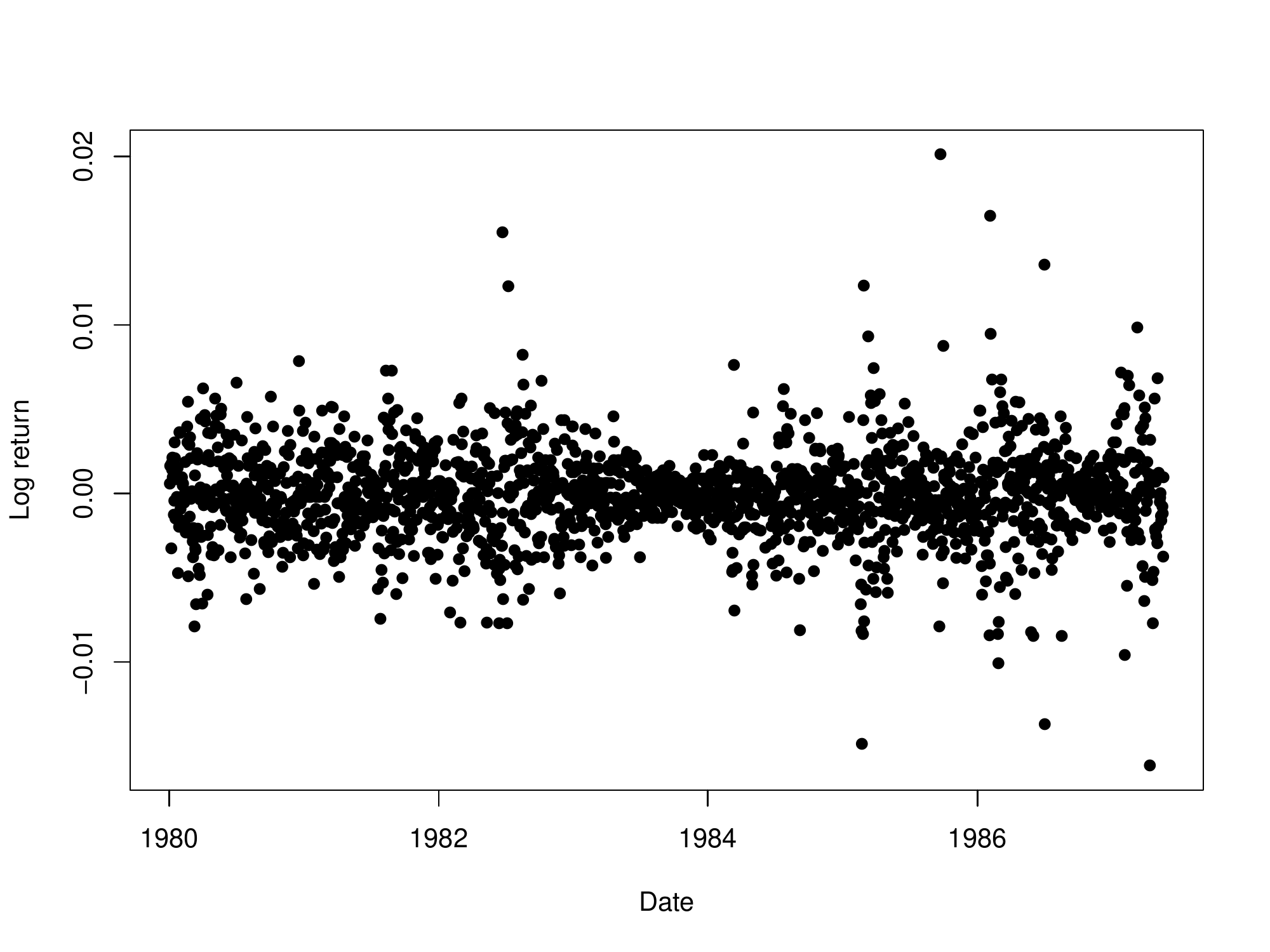}
\caption{Log returns for US dollar / Canadian Dollar exchange rates.}
\label{fig:fx_data}
\end{figure}

\begin{figure}[htp]
\includegraphics[width=\textwidth]{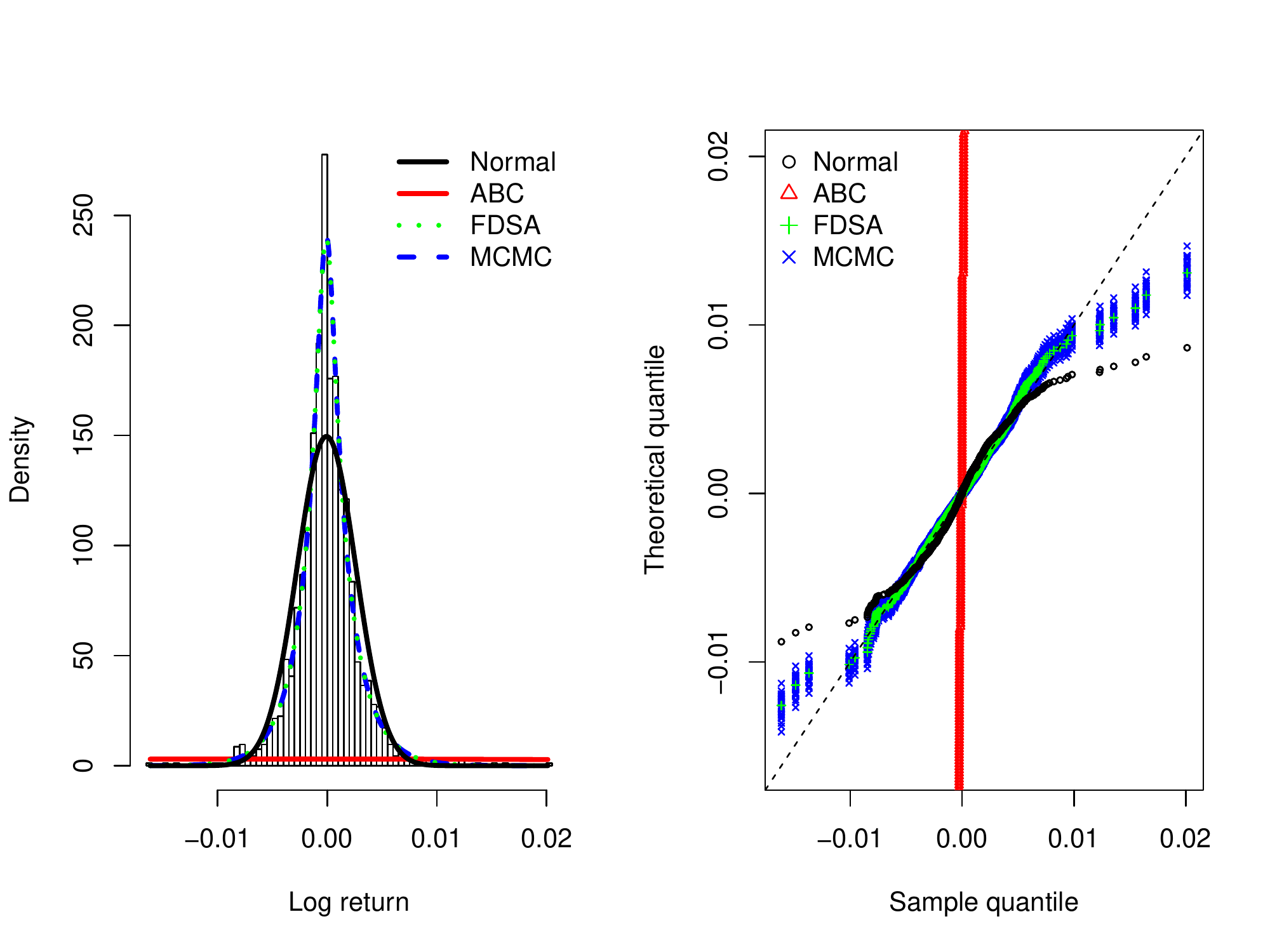}
\caption{(Left) Histogram of exchange rate log returns, and fitted $g$-and-$k$ densities.
(Right) Quantile-quantile (QQ) plots of fitted $g$-and-$k$ densities.
QQ plots are shown for 30 vectors of parameters sampled from the second half of the MCMC output.}
\label{fig:gk_fits}
\end{figure}

We illustrate \pkg{gk}'s inference methods on the \code{Garch} exchange rate dataset from the \pkg{Ecdat} package.
This consists of 1967 daily US dollar exchange rates against other currencies from 1980 to 1987.
We concentrate on the exchange rate with Canadian Dollars.
Let $x_t$ denote the exchange rate on day $t$.
The \dfn{log return} is defined as $\log(x_{t+1}/x_t)$.
Figure \ref{fig:fx_data} is a time series plot of the log returns.
Figure \ref{fig:gk_fits} shows a histogram and a quantile-quantile plot indicating that the tails are heavier than those of a normal density.

We focus on using the $g$-and-$k$ distribution to model the log returns under an IID assumption.
For models also including time series structure see for example \cite{Drovandi:2011}.
The full code for the analysis below can be run via the \code{fx} function.

The ABC and MCMC analyses which follow are Bayesian and require specification of a prior.
We use a uniform prior for ease of comparison to the maximum likelihood results from FDSA.
For MCMC we are able to use an improper uniform prior.
For ABC a proper prior is required so we bound the parameters as follows
$-1<A<1$, $0<B<1$, $-5<g<5$, $0<k<10$.
We restrict $A$ and $B$ to magnitude 1 at most, as we believe log returns of this magnitude are highly unlikely.
The $g$ and $k$ parameters are given wider support which can capture a broad range of distributional shapes.

\paragraph{ABC}

We ran ABC as follows:
\begin{example}
rprior = function(i) {cbind(runif(i,-1,1), runif(i,0,1), runif(i,-5,5), runif(i,0,10))}
abc_out = abc(log_return, N=1E7, rprior=rprior, M=200, sumstats='moment estimates')
\end{example}
This simulated $10^7$ parameter vectors and accepting the best $200$.
We used moment estimator summary statistics, described earlier, which can be simulated quickly without the need to simulate an entire dataset.
As a result this analysis took only 6 minutes.

\begin{figure}[htp]
\includegraphics[width=\textwidth]{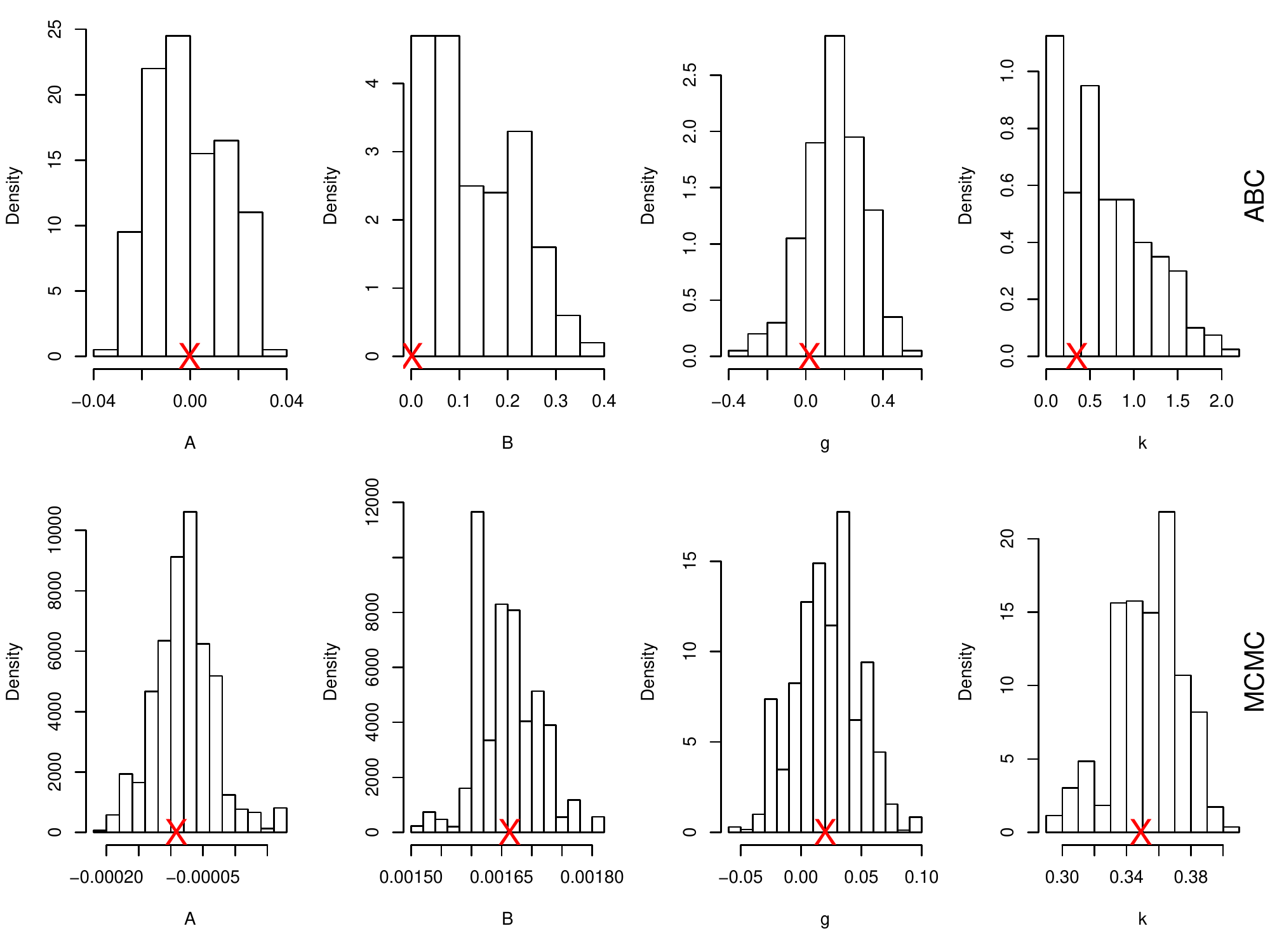}
\caption{Parameter inference for fitting the $g$-and-$k$ distribution to exchange rate log returns.
The top row shows the ABC posterior sample and the bottom row the MCMC posterior sample, which requires much more concentrated parameter scales.
FDSA estimates of the MLEs are shown by crosses on the $x$-axis.}
\label{fig:posteriors}
\end{figure}

The resulting approximate posterior samples are shown in Figure \ref{fig:posteriors}.
Figure \ref{fig:gk_fits} shows density and quantile-quantile plots under the mean parameter values.
These reveal a very poor fit to the data.
However this short ABC analysis does provide reasonable tuning choices for the other methods.

\paragraph{FDSA}

We ran FDSA as follows:
\begin{example}
abc_out_tf = abc_out[,1:4]
abc_out_tf[,2] = log(abc_out_tf[,2])
abc_est_tf = colMeans(abc_out_tf)
fdsa_out_pilot = fdsa(log_return, N=1E4, logB=TRUE, theta0=abc_est_tf, batch_size=100,
                      a0=2E-4)
a0 = c(1E-6, 1E-2, 1E-2, 1E-2)
fdsa_out = fdsa(log_return, N=1E4, logB=TRUE, theta0=abc_est_tf, batch_size=100, a0=a0)
\end{example}
We found that using the original parameterisation caused high variance in our gradient estimates.
This is because the log-likelihood surface becomes extremely steep for $B$ close to 0.
Therefore we reparameterised $B$ to $\log B$.
The initial FDSA state was set to equal the ABC means.
The FDSA steps sizes $a_0$ were tuned by trial-and-error.

\begin{figure}[htp]
\includegraphics[width=\textwidth]{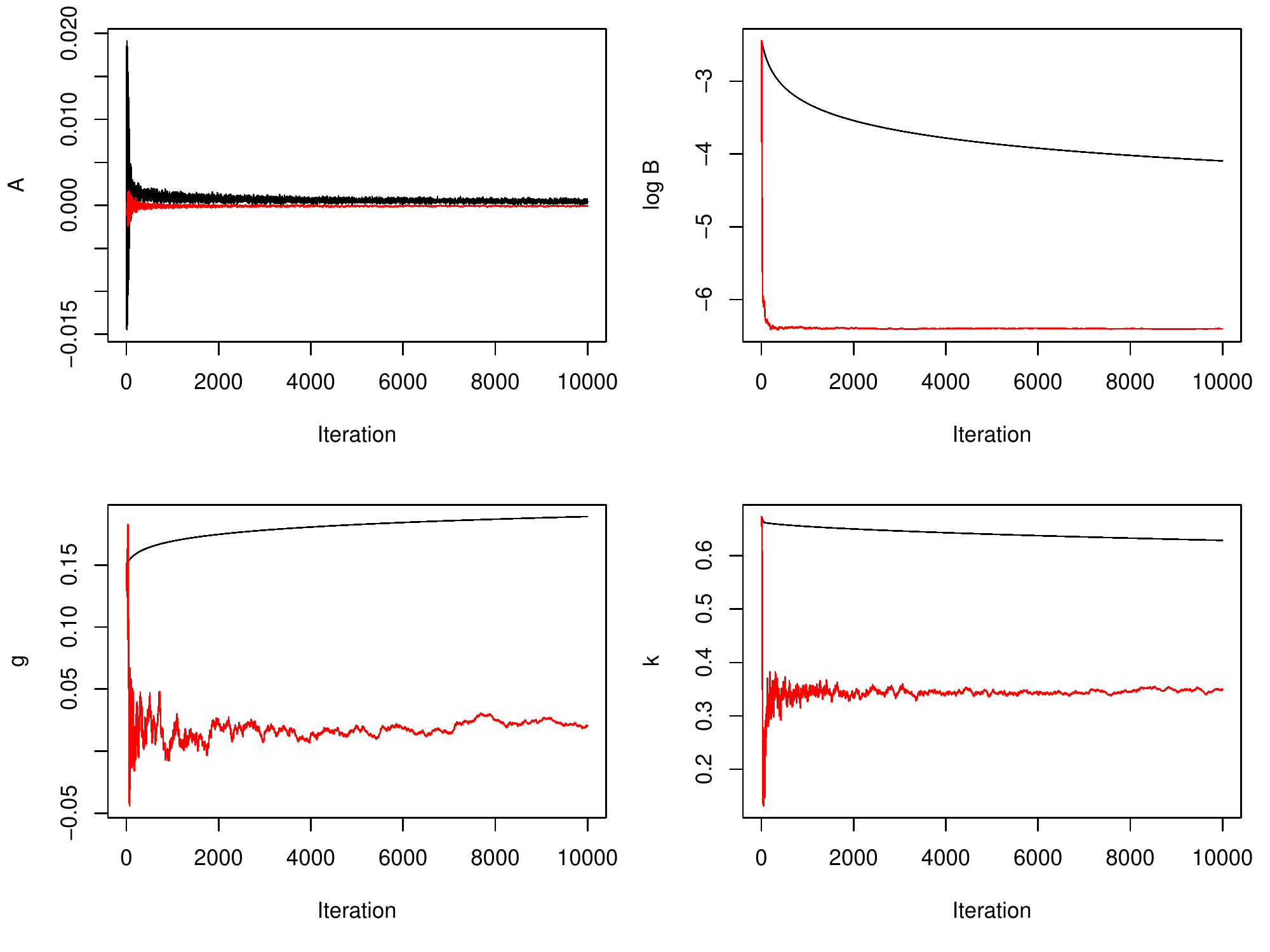}
\caption{Output from the FDSA algorithm to infer $g$-and-$k$ parameters for exchange rate log returns.
Black shows output from a pilot run with $a_0=2 \times 10^{-4}$.
Red shows output from the final run with $a_0=(10^{-6}, 10^{-2}, 10^{-2}, 10^{-2})$.
}
\label{fig:fdsa}
\end{figure}

Figure \ref{fig:fdsa} shows a trace plot of the FDSA algorithm output.
A pilot run with $a_0=2 \times 10^{-4}$ is shown in black.
Parameters $\log B, g$ and $k$ do not converge over $10,000$ iterations.
However they have smooth curves, indicating that there is relatively little noise in their gradient estimates and so larger steps could be taken.
In contrast $A$ converges quickly and then oscillates noisily.
This indicates that a smaller step size could be used to average out this noise more effectively without endangering convergence.
Therefore for the final run we used $a_0=(10^{-6}, 10^{-2}, 10^{-2}, 10^{-2})$.

The final FDSA analysis took 17 minutes.
The final states were $A=9.1 \times 10^{-5}$, $B=1.7 \times 10^{-3}$, $g=2.0 \times 10^{-2}$ and $k=0.35$.
Figure \ref{fig:gk_fits} shows density and quantile-quantile plots under these parameter values.
These are a much better fit to the data than the ABC results.

Next we use the FDSA results to help tune an MCMC algorithm,
which quantifies the uncertainty in the parameter values.

\paragraph{MCMC}

We ran MCMC as follows:
\begin{example}
fdsa_est_tf = fdsa_out[1E5,1:4]
Sigma0 = var(fdsa_out[1E5 + (-1000:0),1:4])
log_prior = function(theta) {
    if (theta[4]<0) return(-Inf)
    return(theta[2])
}
mcmcout_tf = mcmc(log_return, N=1E4, logB=TRUE, get_log_prior=log_prior,
                  theta0=fdsa_est, Sigma0=Sigma0)
\end{example}
Again we used a log reparameterisation for $B$.
To achieve an improper uniform prior on the original parameterisation,
we used a prior density proportional of $B \mathbb{1}(k>0)$ on $(A,\log B, g, k)$
(where $\mathbb{1}$ represents an indicator function).
Our initial parameter vector was the final FDSA state.
We use the variance matrix of the last 1000 FDSA states to select the initial MCMC proposal variance.

\begin{figure}[htp]
\includegraphics[width=\textwidth]{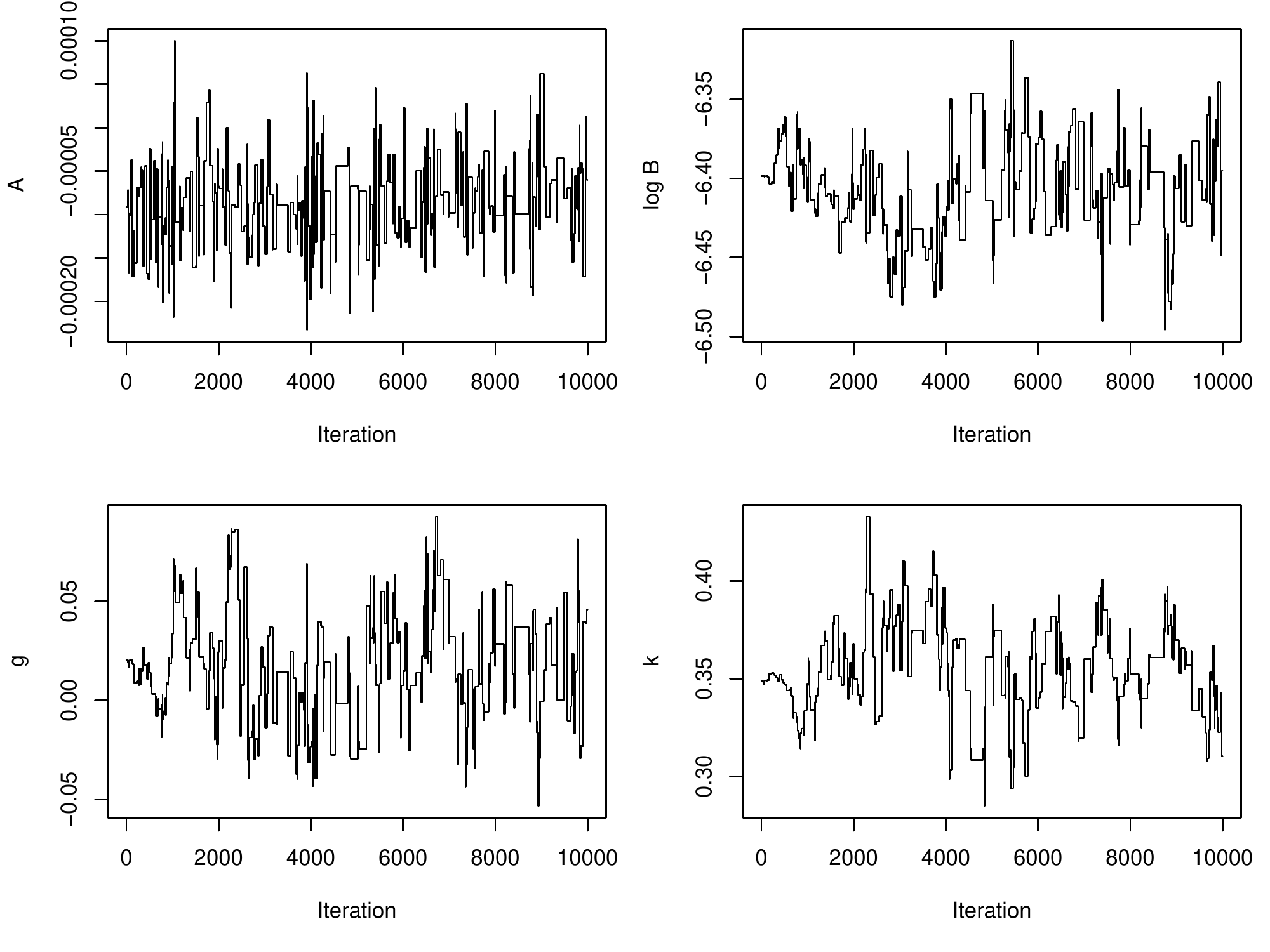}
\caption{States of an MCMC algorithm to infer $g$-and-$k$ parameters for exchange rate log returns.}
\label{fig:mcmc}
\end{figure}

Figure \ref{fig:mcmc} shows a trace plot of the MCMC algorithm output.
For the first few hundred iterations small proposals are made, at least for $\log B$, $g$ and $k$,
but the proposal variance quickly adapts and the remainder of the output appears to have converged.
Exploratory work showed that taking a poor initial state meant MCMC is very slow to converge, because the variance matrix adapts to the transient state of the algorithm.
Hence tuning based on FDSA output is very useful.

The MCMC analysis took 39 minutes.
Figure \ref{fig:posteriors} parameter histograms and figure \ref{fig:gk_fits} shows density and quantile-quantile plots.
These are similar to the FDSA fit.
Note that the density plot is based on mean parameter values from the MCMC output
(after discarding the first half of the output as burn-in
and transforming $\log B$ values back to the original parameterisation).

\paragraph{Summary}

The ABC analysis is quick but produces a poor fit.
However it helps tune the FDSA method
which finds a good estimate of the MLE in a reasonable time.
Further computational effort using MCMC provides a Bayesian fit.
Figure \ref{fig:gk_fits} shows that the $g$-and-$k$ distribution fits the data better than a normal distribution, but still does not fit the most extreme observations.
Further improvements might be possible by using more flexible distributions,
for example allowing different $k$ parameters for the upper and lower tails \citep{Peters:2016}.

\section{Discussion}

This paper has reviewed the $g$-and-$k$ and $g$-and-$h$ distributions,
and introduced the \pkg{gk} package to work with them.
The package includes the usual distributional functions, although the pdf and cdf functions are slow due to relying on numerical root-finding.
Another function tests the validity of different parameter combinations, and this was used to produce a novel result on which parameters are valid for the $g$-and-$k$ distribution
(i.e.~it is appears to be sufficient that $k \geq \max(-0.5, -0.045-0.01g^2)$.)
The package also provides several methods for inference of IID data under these distributions,
and their use has been illustrated above.
The methods include a FPSA algorithm which can find MLEs for large datasets in a reasonable time
and has not been applied to this problem before.

\appendix
\section{Appendix A: Formulae}

The derivatives of the $Q$ functions are as follows:
\begin{align}
Q_{gk}'(z;A,B,g,k,c) &= B (1+z^2)^k R_{gk}(z;g,k), \label{eq:Qgk'}\\
Q_{gh}'(z;A,B,g,h,c) &= B \exp(hz^2/2) R_{gh}(z;g,h) \label{eq:Qgh'},
\end{align}
where
\begin{align}
R_{gk}(z;g,k,c) &= \left[ 1 + c \tanh(gz/2) \right] \frac{1+(2k+1)z^2}{1+z^2} + \frac{cgz}{2 \cosh^2(gz/2)}, \label{eq:Rgk} \\
R_{gh}(z;g,h,c) &= \left[ 1 + c \tanh(gz/2) \right] (1+hz^2) + \frac{cgz}{2 \cosh^2(gz/2)}. \label{eq:Rgh}
\end{align}
Observe that each $Q'$ function has the same sign and roots as the corresponding $R$ function.

\section{Appendix B: Range of valid parameters - theory}

This appendix proves theoretical results quoted earlier about which parameter values produce valid $g$-and-$k$ and $g$-and-$h$ distributions.

First note that the defining functions in \eqref{eq:Qgk} and \eqref{eq:Qgh}
both have the property that $Q(z;A,B,-g,k,c) = Q(z;A,B,g,k,-c)$.
Therefore any behaviour produced by $c<0$ can be replicated with $c>0$ and a different choice of $g$.
So for simplicity it suffices to concentrate on $c \geq 0$.

For the remainder of this appendix, distributional validity will correspond to a strictly increasing quantile function.
This property is generally violated if $c>1$, as there are two solutions to $Q(z)=A$: $z=0$ and a solution to $1+c\tanh(gz/2)=0$ (The only exception is the special case of $g = 0$.)
Also taking $h<0$ or $k<-1/2$ is invalid, as in either case $Q$, which is continuous, has a positive gradient at $z=0$ but limits of zero.

\begin{figure}[htp]
\includegraphics[width=\textwidth]{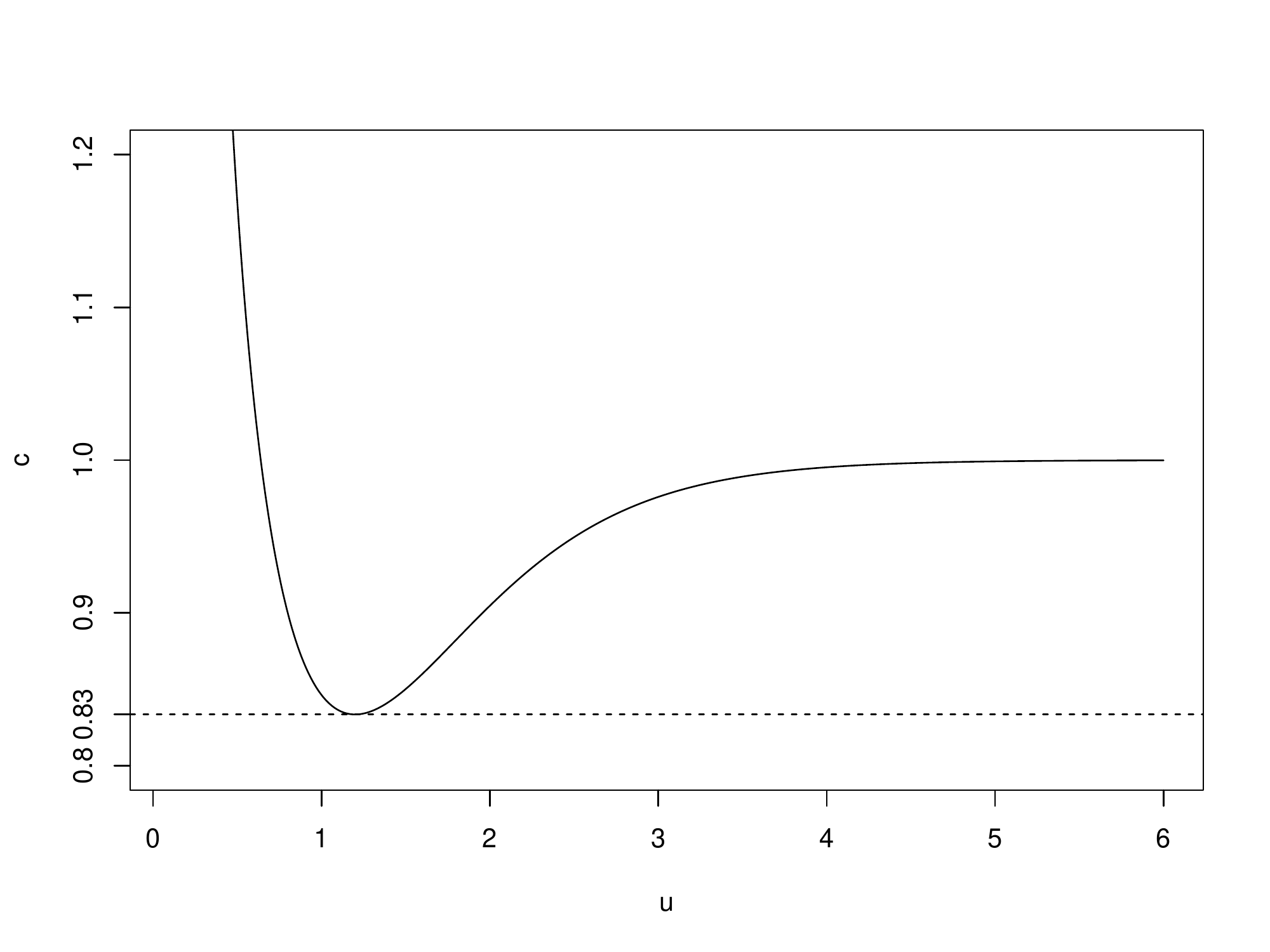}
\caption{Solutions to \eqref{eq:c equality} for $v=1$ and $u>0$.}
\label{fig:cplot}
\end{figure}

Finally it is shown that non-negative values of $k$ or $h$ produce valid distributions provided that $0 \leq c<c^* \approx 0.83$ \citep{Rayner:2002}.
From Appendix A it suffices to derive the values of $c$ such that $R(z)$
-- representing either $R_{gk}(z;g,k,c)$ or $R_{gh}(z;g,h,c)$ --
is guaranteed to be positive for $k \geq 0$ or $h \geq 0$.
Note that $R(z)$ is a continuous function of $z$, and $R(0)>0$.
So a sufficient condition for validity is that no solution to $R(z)=0$ exists.
Rearranging $R(z)=0$ using \eqref{eq:Rgk} and \eqref{eq:Rgh} gives
\begin{align}
1/c &= u v \sech^2 u + \tanh u, \label{eq:c equality} \\
\text{where} \qquad u&=-gz/2, \nonumber \\
\text{and} \qquad v&= 
\begin{cases}
\frac{1+z^2}{1+(2k+1)z^2} & \text{($g$-and-$k$)}\\
\frac{1}{1+hz^2} & \text{($g$-and-$h$)}
\end{cases} \nonumber
\end{align}
For $k \geq 0$ or $h \geq 0$, $v$ can only take values in $(0,1]$ with $1$ attained by $z=0$.
Hence \eqref{eq:c equality} gives $c>0$ if and only if $u>0$, and we concentrate on this case from now on.
We wish to find the minimum positive solution for $c$.
Since $1/c$ is increasing in $v$ it suffices to concentrate on its largest value, $v=1$.
The problem reduces to minimising $(u \sech u + \tanh u)^{-1}$ for $u>0$.
Numerically this gives $c^* \approx 0.83$, as shown in Figure \ref{fig:cplot}.

\section{Acknowledgements}

Thanks to Kieran Peel who wrote a helpful undergraduate dissertation on this topic.

\bibliography{prangle}

\address{Dennis Prangle\\
  Department of Mathematics and Statistics\\
  Newcastle University\\
  NE1 7RU\\
  UK\\}
\email{dennis.prangle@newcastle.ac.uk}
\end{article}

\end{document}